\documentstyle[12pt]{article}

\begin{document}

\begin{center}

{}~\vfill

{\large \bf PROJECTIVE INVARIANCE AND ONE-LOOP \protect \\
{}~\vfill
EFFECTIVE ACTION IN AFFINE-METRIC GRAVITY \protect \\
{}~\vfill
INTERACTING WITH SCALAR FIELD }

\vfill

{\large M. Yu. Kalmykov}
\footnote {Permanent address:
{\em Bogoliubov Laboratory of Theoretical Physics,
 Joint Institute for Nuclear  Research,
 $141~980$ Dubna $($Moscow Region$)$, Russian Federation} \protect \\
 E-mail: $kalmykov@thsun1.jinr.dubna.su$
 \protect \\
Supported in part by ISF grant \# RFL000},
 ~{\large P. I. Pronin}\footnote{E-mail: $pronin@theor.phys.msu.su$}
 and~ {\large K. V. Stepanyantz}

\vspace{2cm}

{\em Department of Theoretical Physics, Physics Faculty\\
 Moscow State University, $117234$, Moscow, Russian  Federation}

\end{center}

\vfill

\begin{abstract}

We investigate the influence of the projective invariance on the
renormalization properties of the theory. One-loop counterterms
are calculated in the most general case of interaction of gravity
with scalar field.

\end{abstract}

\vfill

\pagebreak

\section{Introduction}

The construction of  quantum theory of gravity is an unresolved problem
of modern theoretical physics. It is well know that the Einstein theory
of gravity is not renormalizable  in an ordinary sense
{}~\cite{tHV} --~\cite{vandeven}.  Therefore, one  needs to modify
the theory or to show, that the difficulties presently encountered in
the theory are only artifacts of perturbation theory. The simplest
method of modifying the Einstein theory is to introduce terms quadratic
in the curvature tensor in the action of the theory.

\begin{equation}
{\it L}_{gr}  = \left(-\frac{1}{{\it k}^2}R + \alpha R^2_{\mu \nu } +
\beta  R^2\right) \sqrt{-g}
\end{equation}

This theory is renormalizable and asymptotically  free
 but it is not unitary because the ghosts and tachyons are present in
the spectrum of the theory ~\cite{Stelle} -- ~\cite{Fradkin}.
It should be noted that it is impossible to restore the
unitarity of the theory by means of loop corrections or adding an
interaction with matter fields ~\cite{AT} , ~\cite{J}.
Hence, one needs to use a new method in order to construct the
theory of gravity.

Among  various methods of constructing the quantum theory of
gravity one should emphasize the gauge approach as the most
promising ~\cite{H-76} -- ~\cite{H-85}.
In gauge treatment of gravity there are two sets of dynamical
variables, namely, the vierbein $h^a_{~\mu} (x)$ and local Lorentz
connection $\omega^a_{~b\mu }(x)$ or metric $g_{\mu  \nu }(x)$ and
affine connection $\Gamma^\sigma _{~\mu \nu }(x)$. The theory based on
the first set of variables is called the Poincar\`{e} gauge
gravitational theory with the structure group $P_{10}$ ~\cite{AAT},
{}~\cite{M-93}. A curvature tensor $R^a_{~b \mu  \nu }(\omega )$ and
a torsion tensor $Q^a_{~\mu  \nu }(h,\omega)$, which are the strength
tensors of the Poincar\`{e} gauge gravitational theory, are defined by
the following relations:

\begin{eqnarray}
R^a_{~b \mu \nu }(\omega) & =  & \partial_\mu \omega^a_{~b \nu
} - \partial_\nu \omega^a_{b \mu }  + \omega^a_{~c \mu } \omega^c_{b
\nu } - \omega^a_{~c \nu } \omega^c_{b \mu }   \\
Q^a_{~\mu  \nu }(h,\omega ) & = & -\frac{1}{2}
\left(\partial_\mu h^a_{~\nu } -
\partial_\nu h^a_{~\mu } + \omega^a_{~c \nu } h^c_{~\mu } -
\omega^a_{~c \mu } h^c_{~\nu } \right)
\end{eqnarray}

The theory based on the second set of variables is called the affine
gauge gravitational theory with the structure gauge group
$GA(4,R)$ ~\cite{Mans} -- ~\cite{Neem-88}.
The strength tensor of the theory is the curvature tensor
$\tilde R^\sigma _{~\lambda  \mu \nu }(\Gamma )$ defined as:

\begin{equation}
\tilde R^\sigma _{~\lambda  \mu  \nu }(\Gamma ) = \partial_\mu
\Gamma^\sigma _{~\lambda \nu }  - \partial_\nu \Gamma^\sigma _{\lambda
\mu } + \Gamma^\sigma_{~\alpha \mu } \Gamma^\alpha_{~\lambda  \nu } -
\Gamma^\sigma_{~\alpha  \nu }  \Gamma^\alpha_{\lambda \mu }
\end{equation}
The Lagrangian of a gauge theory is  built out of
terms quadratic in the strength tensor of fields. In the Poincar\`{e} or
affine gauge theories the Lagrangians are defined as :

\begin{eqnarray}
{\it L}_{P_{10}} & = &
\left(\frac{A_i}{{\it k}^2}Q^2(h,\omega ) + B_j R^2(\omega) \right)
\sqrt{-g}  \\
{\it L}_{GA(4,R)} & = &
C_j \tilde R^2(\Gamma ) \sqrt{-g}  \\
\end{eqnarray}
where $A_i, B_j, C_j$ are arbitrary constants and $R^2, Q^2$ are now a
symbolic notation for the contractions of the curvature tensors or the
torsion tensors respectively.

 At the present time there are a lot of papers concerning the classical
problems of these theories ~\cite{HS} -- ~\cite{H-89}.
For example, it is possible to find some coefficients $A_i$ and $B_j$
in the Poincar\`{e} gauge gravitational theory in order to obtain
a unitary model ~\cite{Nev} -- ~\cite{Kun}.
However, the renormalizability properties of the theories have been
studied insufficiently ~\cite{Martel} -- ~\cite{MKL}.

In the affine-metric theory of gravity there are  models possessing an
extra projective symmetry. By the projective invariance we mean that
the action is invariant under the following transformation of fields:

\begin{eqnarray}
x^\mu & \rightarrow &  'x^\mu  = x^\mu \nonumber \\
g_{\mu  \nu }(x) & \rightarrow & 'g_{\mu \nu }(x)  = g_{\mu \nu }(x)
\nonumber \\
\Phi_{mat}(x) & \rightarrow & '\Phi_{mat}(x)  =  \Phi_{mat}(x)
\nonumber \\
\Gamma^\sigma _{~\mu \nu }(x) & \rightarrow  &
'\Gamma^\sigma _{~\mu \nu }(x)  = \Gamma^\sigma _{~\mu \nu }(x)
+ \delta^\sigma_\mu C_\nu(x)
\label{projective}
\end{eqnarray}
where $C_\nu (x)$ is an arbitrary vector.

The classical properties of models with the projective invariance have
been discussed in papers ~\cite{Sand}, ~\cite{HKMM-91}.
However, the quantum properties of
the projective invariance have not been investigated. It should be
noted, that the presence of an additional symmetry in the theory may
improve the renormalization properties of the theory. For example,
because of the presence of supersymmetry, the terms violating the
renormalizability of supergravity, show up only in higher loops.
So, the projective invariance may have the considerable role for the
renormalizability of the theory. In order to investigate the influence
of the projective invariance on renormalizability of the theory one
needs to calculate the counterterms in some model possessing the
projective invariance. The simplest model of this type is the
model with the Lagrangian:

\begin{equation}
{\it L}_{gr}  = -\frac{1}{{\it k}^2} \tilde R(\Gamma) \sqrt{-g}
\label{simple}
\end{equation}

But because of the  degeneracy of the four-dimensional
space-time~\cite{PVN}, the terms violating the renormalizability of the
theory arise only at two-loop level. The two-loop  calculations are
very cumbersome.  Since we would like to restrict ourselves to the
one-loop calculations and to investigate the influence of the
projective invariance on the renormalizability of the models, we
consider the interaction of the gravity with a matter field.

\begin{equation}
{\it L}_{gr}  = \Biggl( \left(\xi \varphi^2 - \frac{1}{{\it
k}^2}\right) \tilde R(\Gamma)  + \frac{2}{{\it k}^2}
\Lambda  + \frac{1}{2}\partial_\mu \varphi \partial_\nu
\varphi g^{\mu \nu } \Biggr) \sqrt{-g}
\label{model}
\end{equation}
where $\Lambda $ is a cosmological constant.

\pagebreak

We consider $\Gamma^\sigma _{~\mu \nu }(x), \phi(x), g_{\mu \nu }(x)$
as independent dynamical fields. This model is invariant
under the projective transformation (\ref{projective}) and general
coordinate transformation:

\begin{eqnarray}
x^\mu & \rightarrow &  'x^\mu  = x^\mu + \xi^\mu (x) \nonumber \\
g_{\mu  \nu }(x) & \rightarrow & 'g_{\mu \nu }(x)  = g_{\mu \nu }(x)
- \partial_\mu \xi^\alpha  g_{\alpha \nu }(x)
- \partial_\nu \xi^\alpha  g_{\alpha \mu }(x)
- \xi^\alpha \partial_\alpha  g_{\mu \nu }(x)
\nonumber \\
\varphi (x) & \rightarrow & '\varphi(x)  =  \varphi(x)
- \xi^\alpha \partial_\alpha  \varphi (x)
\nonumber \\
\Gamma^\sigma _{~\mu \nu }(x) & \rightarrow  &
'\Gamma^\sigma _{~\mu \nu }(x)  = \Gamma^\sigma _{~\mu \nu }(x)
- \partial_\mu \xi^\alpha  \Gamma^\sigma_{~\alpha \nu }(x)
- \partial_\nu \xi^\alpha  \Gamma^\sigma_{~\mu \alpha }(x)
\nonumber \\
&& ~~~~~~~~~~~ + \partial_\alpha   \xi^\sigma   \Gamma^\alpha_{~\mu
\nu }(x) - \xi^\alpha \partial_\alpha  \Gamma^\sigma_{~\mu \nu }(x) -
\partial_{\mu \nu } \xi^\sigma
\label{coordinate}
\end{eqnarray}

The main aim of our paper is to research the influence of
the projective invariance on the renormalization properties of
the theory.  In particular, we consider the following problems
in the next section:

\begin{enumerate}
\item A necessity of introducing the term fixing
the projective invariance at the quantum level.
\item  The presence of the ghosts
connected with the projective invariance.
\item The addition of the "projective" ghosts
to the one-loop effective action
\end{enumerate}

We use the following notations:
$$ c = \hbar = 1;~~~~~ \mu , \nu  = 0,1,2,3;~~~~~ {\it k}^2 = 16 \pi G$$
$$ \tilde R_{\mu \nu }(\Gamma ) =
\tilde R^\sigma_{~\mu \sigma \nu }(\Gamma ),~~~~~
 \tilde R(\Gamma ) = \tilde R_{~\mu \nu }(\Gamma ) g^{\mu \nu },
{}~~~~~(-g) = det(g_{\mu \nu }) $$

The objects marked by the tilde $\tilde{} $  are constructed by means of
the affine connection $\Gamma^\sigma_{~\mu \nu }$. The others are the
Riemannian objects.

\section{One-loop counterterms}

For calculating the one-loop effective action we use the background
field method ~\cite{BDW-67}, ~\cite{BDW}. In accordance with this
method all dynamical variables are rewritten as the sum of the classical
and quantum parts. In general case, the dynamical variables in the
affine-metric theory are $\Gamma^\sigma_{\mu \nu },
\bar g_{\mu \nu } = g_{\mu \nu }(-g)^r, \bar \varphi  = \varphi (-g)^s $,
where $r,s$ are arbitrary numbers satisfying the only condition:
$r \neq -\frac{1}{4}$. The one-loop counterterms on the mass-shell do
not depend on the value of $r$ and $s$. To simplify our  calculation we
use the following values: $r=s=0$.

The fields $\Gamma^\sigma_{~\mu \nu } ,g_{\mu \nu },\varphi $ are now
rewritten according to

\begin{eqnarray}
\Gamma^\sigma_{~\mu \nu } & = & \Gamma^\sigma_{\mu \nu }
+ {\it k} \gamma^\sigma_{~\mu \nu }  \nonumber \\
g_{\mu \nu }  & = & g_{\mu \nu } + {\it k}h_{\mu \nu } \nonumber \\
\varphi  & = &  \frac{1}{{\it k}} \varphi + \phi
\label{expansion}
\end{eqnarray}
where $\Gamma^\sigma_{~\mu \nu } ,g_{\mu \nu },\varphi $
are the classical parts satisfying the following equations

\begin{eqnarray}
\frac{\delta {\it S}}{\delta \Gamma^\sigma_{~\mu \nu }} = 0
& \Rightarrow & D^\sigma_{~\mu \nu } = - \frac{1}{2}
\frac{1}{\alpha(\varphi)} \partial_\lambda \alpha(\varphi )
(g^{\lambda \sigma }g_{\mu \nu } - \delta^\lambda_\mu \delta^\sigma_\nu  ) +
\delta^\sigma_\mu C_\nu
\nonumber \\
\frac{\delta {\it S}}{\delta g_{\mu \nu }} = 0
& \Rightarrow & -\alpha(\varphi) \tilde R_{(\mu \nu) }(\Gamma )  =
\frac{1}{2} \partial_\mu \varphi \partial_\nu \varphi  + \Lambda g_{\mu
\nu }
\nonumber \\
\frac{\delta {\it S}}{\delta \varphi } = 0 &
\Rightarrow & 2\xi \varphi \tilde R(\Gamma )  - g^{\mu \nu } \nabla_\mu
\nabla_\nu \varphi  = 0
\label{mass-shell}
\end{eqnarray}
where
\begin{eqnarray} D^\sigma_{~\mu \nu } & = & \Gamma^\sigma_{~\mu \nu } -
g^{\sigma \lambda } \frac{1}{2} \left(- \partial_\lambda g_{\mu \nu } +
\partial_\mu  g_{\nu \lambda } + \partial_\nu  g_{\mu \lambda }\right)
\nonumber \\
\alpha (\varphi ) & = & \xi \varphi ^2 - 1  \nonumber
\end{eqnarray}
$C_\nu $ is an arbitrary vector.

The action (\ref{model}) expanded as a power series in the quantum
fields (\ref{expansion}) defines the effective action for calculating
the loop counterterms. The one-loop effective Lagrangian quadratic
in the quantum fields is:

\begin{eqnarray}
{\it L}_{eff} & = & \Biggl( \alpha(\varphi) \frac{1}{2}
\gamma^\sigma_{~\mu  \nu}
F_{\sigma ~~\lambda}^{~\mu \nu ~\alpha \beta}
\gamma^\lambda_{~ \alpha \beta} +
\alpha(\varphi) \frac{1}{2} h_{\mu \nu} h_{\sigma \lambda}
D^{\mu \nu \sigma \lambda}
-  \frac{1}{4} h_{\alpha \beta} h_{\mu \nu}
\Lambda P^{-1 \alpha \beta \mu \nu }
\nonumber \\
& + &  \xi \phi ^2 R
+ \frac{1}{2} g^{\mu \nu} \nabla_\mu \phi  \nabla_\nu \phi
- \frac{1}{2} \alpha(\varphi) h_{\alpha \beta}
P^{-1 \alpha \beta \mu \nu} \biggl(
B_{\lambda~~~\mu \nu}^{~\epsilon \tau \sigma} \nabla_\sigma +
\triangle^{~\epsilon \tau}_{\lambda ~~\mu \nu}
\biggr) \gamma^\lambda_{~ \epsilon \tau}
\nonumber \\
& +  & 2\xi \varphi \phi \Biggl( \biggl(
B_\lambda ^{~\epsilon \tau \sigma} \nabla_\sigma +
\triangle_\lambda^{~\epsilon \tau} \biggr)
\gamma^\lambda _{~\epsilon \tau}
- \frac{1}{2} h_{\alpha \beta} P^{-1 \alpha \beta \mu \nu} R_{\mu
\nu} \Biggr)
\nonumber \\
& + & \frac{1}{2} \nabla_\mu \varphi \nabla_\nu \varphi \biggl(
h^{\mu \lambda}h^\nu_{~\lambda} -
\frac{1}{2}hh^{\mu \nu} - \frac{1}{8}
 h_{\alpha \beta}h_{\sigma \lambda} g^{\mu \nu }
P^{-1 \alpha \beta \sigma \lambda} \biggr)
\nonumber \\
& - & \frac{1}{2}\nabla_\mu \varphi \nabla_\nu \phi h_{\alpha \beta}
P^{-1 \alpha \beta \mu \nu} \Biggr)  \sqrt{-g}
\label{effective}
\end{eqnarray}

where

\begin{eqnarray}
P^{-1 \alpha \beta \mu \nu } & = &
g^{\alpha \mu } g^{\beta \nu } + g^{\alpha \nu }g^{\beta \mu } -
g^{\alpha \beta }g^{\mu \nu }
\nonumber \\
\triangle_{\lambda ~~\mu \nu}^{~ \alpha \beta} & \equiv &
D^\alpha_{~\mu \nu} \delta^\beta_\lambda + D_\lambda
\delta^\alpha_\mu \delta^\beta_\nu - D^\alpha_{~\mu \lambda}
\sigma^\beta_\nu - D^\beta_{~\lambda \nu} \delta^\alpha_\mu
\nonumber \\
B_{\lambda~~~\mu \nu}^{~\alpha \beta \sigma} & = &
\delta^\sigma_\lambda \delta^\alpha_\mu \delta^\beta_\nu
- \delta^\beta_\lambda \delta^\alpha_\mu \delta^\sigma_\nu
\nonumber \\
F_{\alpha ~~\mu}^{~\beta \lambda~\nu \sigma} & = &
g^{\beta \lambda} \delta^\nu_\alpha \delta^\sigma_\mu
- g^{\beta \sigma} \delta^\nu_\alpha \delta^\lambda_\mu
+ g^{\nu \sigma} \delta^\lambda_\alpha \delta^\beta_\mu
- g^{\lambda \nu} \delta^\sigma_\alpha \delta^\beta_\mu
\nonumber \\
\triangle_{\lambda}^{~\epsilon \tau} & = &
\triangle_{\lambda~~\mu \nu }^{~\epsilon \tau} g^{\mu \nu }
\nonumber \\
B_\lambda^{~\epsilon \tau \sigma} & = &
B_{\lambda ~~~\mu \nu }^{~\alpha \beta \sigma} g^{\mu \nu }
\nonumber \\
D^{\alpha \beta \mu \nu} & = & 2R^{\alpha \mu} g^{\beta \nu} -
RP^{\alpha \beta \mu \nu} - R^{\alpha \beta} g^{\mu \nu}
\end{eqnarray}

Now we may define the propagators of the quantum fields
$\gamma^\sigma_{~\mu \nu}, h_{\mu \nu }, \phi$.
The propagator of the quantum field $\gamma^\sigma_{~\mu
\nu}$  satisfies two conditions:

\begin{equation}
F^{-1 \sigma ~~\lambda}_{~~~~\mu \nu ~\alpha \beta} =
F^{-1 \lambda ~~\sigma}_{~~~~\alpha \beta ~\mu \nu}
\label{first}
\end{equation}

\begin{equation}
F^{-1 \sigma ~~\lambda}_{~~~~\mu \nu ~\alpha \beta}
F_{\lambda~~\rho}^{~\alpha \beta ~ \tau \epsilon} =
\delta^\sigma_\rho \delta^\tau_\mu \delta^\epsilon_\nu
\label{second}
\end{equation}

However, because of the projective invariance of the effective
Lagrangian (\ref{effective}) the propagator does not exist. Under
transformation (\ref{projective}) the quantum part of the
connection transforms as

\begin{equation}
\gamma^\sigma_{~\mu \nu}(x) \rightarrow
'\gamma^\sigma_{~\mu \nu}(x) = \gamma^\sigma_{~\mu \nu}(x)
+ \delta^\sigma_\mu C_\nu(x)
\end{equation}

In order to fix the projective invariance we use the following
condition:

\begin{equation}
f_\lambda = \biggl(
B_1g_{\lambda \sigma} g^{\alpha \beta} +
B_2\delta^\alpha_\sigma \delta^\beta_\lambda +
B_3 \delta^\beta_\sigma \delta^\alpha_\lambda
\biggr) \gamma^\sigma_{~\alpha \beta} \equiv
f_{\lambda  \sigma }^{~~~\alpha  \beta }
\gamma^\sigma_{~\alpha  \beta }
\label{prgf}
\end{equation}

\begin{equation}
{\it L}_{gf}  = \frac{1}{2} f_\mu f^\mu
\end{equation}

where $B_j$ are the constants satisfying the only condition:

\begin{equation}
B_1 + B_3 + 4B_2 \neq 0
\end{equation}

  The action of the projective ghosts defined by the standard
way has the following structure:

\begin{equation}
{\it L}_{gh}  = \overline{\chi}^\mu g_{\mu \nu} (-g)^\alpha \chi^\nu
\label{ghost}
\end{equation}
where

$\overline{\chi}^\mu , \chi^\nu  $ are the grassmann variables;
{}~$\alpha $ is a constant.

The one-loop contribution of the projective ghosts to the effective
action is proportional to the $\delta^4(0)$. In the dimensional
regularization $[\delta^4(0)]_R = 0$ and the contribution of the
projective ghosts to the one-loop counterterms is equal to zero.

Now, we must change the equation (\ref{second}).
The propagator of the quantum field $\gamma^\sigma_{~\mu \nu }$
satisfies equation (\ref{first}) and new condition:

\begin{equation}
F^{-1 \sigma ~~\lambda}_{~~~~\mu \nu ~\alpha \beta}
\overline F_{\lambda~~\rho}^{~\alpha \beta ~ \tau \epsilon} =
\delta^\sigma_\rho \delta^\tau_\mu \delta^\epsilon_\nu
\label{instead}
\end{equation}

where

\begin{eqnarray}
\overline F_{\sigma ~~\lambda}^{~\alpha \beta ~\mu \nu} & = &
F_{\sigma ~~\lambda}^{~\alpha \beta ~\mu \nu} +
f_{\tau \sigma }^{~~~\alpha \beta } f^{\tau ~~\mu \nu }_{~\lambda }
\nonumber \\
& = &
g^{\mu \nu} \delta^\alpha_\lambda \delta^\beta_\sigma (1 + B_1
B_3) + g^{\alpha \beta} \delta^\mu_\sigma \delta^\nu_\lambda (1 + B_1
B_3) - g^{\nu \alpha} \delta^\mu_\sigma \delta^\beta_\lambda - g^{\mu
 \beta}\delta^\nu_\sigma \delta^\alpha_\lambda
\nonumber \\
& + & B_1 B_2
g^{\alpha \beta}\delta^\nu_\sigma \delta^\mu_\lambda +
B_1 B_2
g^{\mu \nu}\delta^\alpha_\sigma \delta^\beta_\lambda +
B_1^2~ g_{\sigma \lambda} g_{\mu \nu} g_{\alpha \beta} +
B_3^2~ g^{\alpha \mu} \delta^\nu_\lambda \delta^\beta_\sigma +
\nonumber \\
& + &  B_2 B_3
g^{\mu \beta} \delta^\alpha_\sigma \delta^\nu_\lambda +
B_2 B_3
g^{\alpha \nu} \delta^\beta_\sigma \delta^\mu_\lambda +
B_2^2~ g^{\nu \beta}\delta^\alpha_\sigma \delta^\mu_\lambda
\end{eqnarray}

Having solved equations (\ref{first}),(\ref{instead})  we obtain the
following result:

\begin{eqnarray}
F^{-1 \alpha ~~\mu}_{~~~~\beta \sigma ~\nu\lambda} & = &
 -\frac{1}{4}  g^{\alpha \mu} g_{\beta \sigma} g_{\nu \lambda} +
  \frac{1}{2}  g^{\alpha \mu} g_{\beta \nu} g_{\sigma \lambda}
- \frac{1}{4} g_{\nu \beta} \delta^\mu_\lambda \delta^\alpha_\sigma
\nonumber \\
& + & \frac{1}{4}  \biggl( g_{\nu \lambda} \delta^\mu_\beta
\delta^\alpha_\sigma
+ g_{\beta \sigma} \delta^\alpha_\nu \delta^\mu_\lambda \biggr)
 - \frac{1}{2} \biggl( g_{\nu \sigma}
\delta^\alpha_\lambda \delta^\mu_\beta + g_{\beta \lambda}
\delta^\mu_\sigma \delta^\alpha_\nu \biggr)
\nonumber \\
& + &
\frac{1}{4} \left( \frac{B_1 - B_3 + 2B_2}{B_1 + B_3 + 4B_2} \right)
\biggl( g_{\nu \lambda}
\delta^\mu_\sigma \delta^\alpha_\beta + g_{\beta \sigma}
\delta^\alpha_\lambda \delta^\mu_\nu \biggr) \nonumber \\
\nonumber \\
& + &
\frac{1}{4} \left( \frac{B_3 - B_1 + 2B_2}{B_1 + B_3 + 4B_2} \right)
\biggl( g_{\beta \lambda}
\delta^\mu_\nu \delta^\alpha_\sigma + g_{\beta \nu} \delta^\alpha_\beta
\delta^\mu_\lambda \biggr) \nonumber \\
\nonumber \\
& + &
\frac{1}{4} \frac{1}{(B_1 + B_3 + 4B_2)^2}
\bigl(4 - B_1^2 - B_3^2 - 12B_2^2 +
\nonumber \\
& + & 10B_1B_3 - 4B_1B_2 - 4B_2B_3 \bigr)
g_{\sigma
\lambda} \delta^\mu_\nu \delta^\alpha_\beta
\end{eqnarray}

To get the diagonal form of the effective Lagrangian we are to replace
the dynamical variables in the following way:

\begin{eqnarray} \gamma^\sigma_{~\mu \nu} \rightarrow \tilde \gamma
^\sigma_{~\mu \nu} = \gamma^\sigma _{~\mu \nu } +
\frac{1}{2} F^{-1 \sigma ~~\lambda}_{~~~~\mu \nu ~ \alpha \beta}
\biggl(B_{\lambda ~~~\rho \epsilon}^{~ \alpha \beta \tau} \nabla_\tau -
\triangle_{\lambda ~~ \rho \epsilon}^{~\alpha \beta}  \biggr)
P^{-1 \rho \epsilon \kappa \upsilon} h_{\kappa \upsilon}
\nonumber \\
+ \frac{1}{2} F^{-1\sigma ~~\lambda}_{~~~~\mu \nu ~ \alpha \beta}
P^{-1 \rho \epsilon \kappa \upsilon} h_{\kappa \upsilon}
B^{~\alpha \beta \eta}_{\lambda ~~~\rho \epsilon}
\frac{1}{\alpha(\varphi)} \nabla_\eta \alpha(\varphi)
\nonumber \\
+ 2 \xi  \phi \varphi  \frac{1}{\alpha(\varphi)}
F^{-1 \sigma~~ \lambda}_{~~~~\mu \nu ~ \alpha \beta}
\triangle_{\lambda}^{~ \alpha \beta} -
\frac{2 \xi}{\alpha(\varphi)} F^{-1\sigma ~~\lambda}_{~~~~\mu \nu ~
\alpha \beta} B^{~\alpha \beta \tau}_{\lambda} \nabla_\tau (\phi
\varphi )
\end{eqnarray}

This replacement does not change the functional measure:

\begin{equation}
{\it det} \frac{\partial \tilde \gamma}{\partial  \gamma} = 1
\end{equation}

We don't give the details of the cumbersome one-loop
calculations that have been performed by means of the
special REDUCE package program created by K.V.Stepanyantz.  One should
note, that we violate  the invariance of the action (\ref{effective})
under the general coordinate transformation by means of the following
gauge ~\cite{Ichi}:

\begin{equation}
F_\mu = \nabla_\nu h_\mu^{~\nu} -
\frac{1}{2} \nabla_\mu h - \frac{2 \xi \varphi}{\alpha(\varphi)}
\nabla_\mu \phi
\end{equation}

\begin{equation}
{\it L}_{gf} = \frac{1}{2} F_\mu F^\mu
\end{equation}

The action of the coordinate ghost is

\begin{equation}
{\it L}_{gh} = \overline c^\mu \biggl(
g_{\mu \nu } \nabla^2 + R_{\mu \nu } - \frac{2 \xi
\varphi}{\alpha(\varphi)} (\nabla_\nu\varphi) \nabla_\mu - \frac{2 \xi
\varphi}{\alpha(\varphi)}(\nabla_\mu \nabla_\nu \varphi) \biggr) c^\nu
\end{equation}

The one-loop counterterms on the mass-shell including the contributions
of the quantum and ghost fields are

\begin{eqnarray}
\triangle \Gamma^1_\infty = - \frac{1}{32 \pi^2 \varepsilon}
\int d^4x \sqrt{-g} \Biggl(
\frac{71}{60} \biggl(
R_{\alpha \beta \mu \nu}R^{\alpha \beta \mu \nu} -
4 R_{\mu \nu } R^{\mu  \nu} + R^2 \biggr)
 \nonumber \\
 + \frac{203}{40} R^2
 + \frac{\Lambda^2}{\alpha^2(\varphi)} \biggl(\frac{463}{5} +
 52 \xi^2 \biggr)
 \nonumber \\
 + \Lambda R
 \biggl( \frac{1}{\alpha(\varphi)}
 \biggl(5 \xi^2 - \frac{4}{3} \xi  + \frac{463}{10} \biggr)
 +
  \frac{\xi^2 \varphi^2}{\alpha^2(\varphi)}
 \biggl(75 \xi + \frac{20}{3} \biggr)  -
 \frac{700}{3}\frac{\xi^4 \varphi^4}{\alpha^3(\varphi)} \biggr)
\Biggr)
\end{eqnarray}

\section{Conclusion}

In our paper we have investigated the influence of the
projective invariance on the renormalizability of the theory. It turns
out that:

\begin{enumerate}
\item In order to define the propagator of the quantum fields
$\gamma^\sigma_{~\mu \nu } $  one needs to fix the projective invariance.
\item The gauge fixing term (\ref{prgf}) has the algebraic structure,
that is it does not contain  derivatives of the fields.
\item The action of the projective ghosts (\ref{ghost}) has also the
algebraic structure. The one-loop contribution of the projective ghosts  is
proportional to the $\delta^4(0)$ Hence its contribution is equal to
zero in the dimensional regularization.
\item The theory involved is not renormalizable. The term
violating the renormalizability of the theory is equal to the $R^2$. It
is easy to show ~\cite{tHV}, that the expression $ \int d^4x \sqrt{-g}
\biggl(R_{\alpha \beta \mu \nu}R^{\alpha \beta \mu \nu} -
4 R_{\mu \nu } R^{\mu  \nu} + R^2 \biggr)
$
is equal to  the $\int d^4x \partial_\mu  j^\mu $. Hence, we can
neglect the contribution of this term to the one-loop counterterms
in space-time without boundaries.

\item The renormalizability of the theory is not affected by the
presence of the projective invariance.
\end{enumerate}

We are greatly indebted to L.V.Avdeev (JINR,Dubna) and our colleagues
>from Department of Theoretical Physics (Moscow State University) for
valuable discussions and suggestions and for  critical reading of
the manuscript.

\end{document}